\documentstyle[12pt]{article}
\let\LARGE=\Large
\let\Large=\large
\let\large=\normalsize
\newcommand{\be}[3]{\begin{equation}  \label{#1#2#3}}     
\newcommand{\ee}{ \end{equation}}
\newcommand{\ba}{\begin{array}}
\newcommand{\ea}{\end{array}}

\def\beq{\begin{equation}}
\def\eeq{\end{equation}}
\def\beqa{\begin{eqnarray}}
\def\eeqa{\end{eqnarray}}

\begin{document}

\begin{titlepage}
\begin{center}
\hfill THU-98/25\\
\hfill {\tt hep-th/9806036}\\

\vskip 3cm

{ \LARGE \bf Dual Heterotic Black-Holes in Four and Two 

Dimensions}

\vskip .3in

{\bf Gabriel Lopes Cardoso$^1$\footnote{\mbox{
\tt 
cardoso@phys.uu.nl}}
and Thomas Mohaupt$^2$\footnote{\mbox{
\tt 
mohaupt@hera1.physik.uni-halle.de}}}
\\

\vskip 1cm

{\em 
\centerline{$^1$Institute for Theoretical Physics, Utrecht University,
3508 TA Utrecht, Netherlands}

\centerline{$^2$Martin-Luther-Universit\"at Halle-Wittenberg, 
Fachbereich Physik,
D-06099 Halle, Germany}}

\vskip .1in

\end{center}

\vskip .2in

\begin{center} {\bf ABSTRACT } \end{center}

We consider a class of extremal and non-extremal 
four-dimensional black-hole solutions
occuring in toroidally compactified heterotic string theory,
whose ten-dimensional interpretation involves a Kaluza-Klein
monopole and a five-brane.  We show that these four-dimensional
solutions can be connected to extremal and non-extremal 
two-dimensional heterotic  black-hole solutions through a change in
the asymptotic behaviour of the harmonic
functions associated with the Kaluza-Klein monopole
and with the five-brane.  
This change in the asymptotic behaviour 
can be achieved by a 
sequence
of $S$ and $T$-$S$-$T$ duality transformations in four 
dimensions. These transformations are implemented 
by performing a 
reduction on a two-torus with Lorentzian 
signature.  
We argue that the same mechanism can be applied to extremal 
and non-extremal
black-hole solutions in the FHSV model.

\vspace{3.5cm}

June 1998\\
\end{titlepage}
\vfill
\eject

\newpage

Some time ago Hyun \cite{hyun} made the important observation
that five-dimensional
black-holes occuring in toroidally compactified type II string 
theory are related by U-duality to two-dimensional black holes.
More recently 
it was shown that the associated entropies agree \cite{SfeSke,teo}.
In \cite{car}
a particular class of extremal four-dimensional black-holes occuring
in toroidally compactified heterotic string theory was connected
to the class of extremal two-dimensional heterotic black-holes 
presented in \cite{guinayo}, and the matching of the associated
entropies was demonstrated.  This connection came about by dropping the
constants appearing in the harmonic functions associated with the
magnetic charges carried by the four-dimensional black-holes, thereby
changing the asymptotic behaviour of the magnetic harmonic functions.
Here, we will extend the results of \cite{car}
to the case of non-extremal black-holes.  We will furthermore
show that the mechanism reported in \cite{bakas}
for changing the asymptotic behaviour of gravitational instantons
can also be used to change the 
asymptotic behaviour of the magnetic harmonic functions.
That is, we will show that there is
a sequence of $S$ and $T$-$S$-$T$ duality
transformations in four dimensions
which achieves that.
This sequence of transformations can be implemented
by considering a reduction on a two-torus with Lorentzian signature.
The four-dimensional heterotic
black-hole solutions under consideration
are thus related by duality
to two-dimensional heterotic black-hole solutions.  We will also argue
that the same applies to solutions occuring in certain $D=4,N=2$ 
string
compactifications, in which the heterotic $N=2$ tree-level 
prepotential does
not receive any perturbative or non-perturbative  corrections.  An example of
such a compactification is provided by 
the FHSV model \cite{fhsv,hm}.

We begin by considering a particular class of four-dimensional 
black-hole solutions occuring in toroidal compactifications of
heterotic string theory.  The solutions under consideration 
are the ones associated with a non-extremal intersection of a string,
a five-brane, a Kaluza-Klein monopole
and a wave in ten dimensions.  The rules for writing down the line element
associated with a non-extremal intersection of a string, a five-brane
and a wave can be found in \cite{cvts}.  A non-extremal version of a 
Kaluza-Klein monopole solution can be obtained by performing a Buscher duality
transformation of a non-extremal five-brane solution.  The Buscher duality
transformation rules state that if the dualization is performed with
respect to an
isometry direction $z$, then in the string frame
the dual metric, 
the dual antisymmetric tensor field and the dual dilaton field
are given by \cite{buscher}
\beqa
{\tilde G}_{zz} &=& \frac{1}{G_{zz}} \;\;\;,\;\;\;
{\tilde G}_{zi} = \frac{B_{zi}}{G_{zz}} \;\;\;,\;\;\;
{\tilde G}_{ij} = G_{ij} - \frac{G_{zi} G_{zj} - B_{zi} B_{zj}}{G_{zz}}
\nonumber\\
{\tilde B}_{zi} &=& \frac{G_{zi}}{G_{zz}} \;\;\;,\;\;\;
{\tilde B}_{ij} = B_{ij} + \frac{G_{zi} B_{zj} - G_{zj} B_{zi}}{G_{zz}}
\;\;\;,\;\;\;
{\tilde \phi} = \phi - \frac{1}{2} \log G_{zz} \;\;\;.\;\;\;
\label{buscher}
\eeqa
Let us then consider the non-extremal five-brane solution in
ten dimensions, whose line element 
in the string-frame is described by \cite{cvts}
\beqa
ds^2_{10} &=&  \left( 
- f  dt^2 + dx_1^2 + \dots + dx_5^2 \right)
+
H_5 \left(  f^{-1} dr^2  +   r^2 \sin^2 \theta d \phi^2 + r^2 d \theta^2
+ dz^2 \right) \;.
\label{nonfiveb}
\eeqa
Here we 
take the harmonic function $H_5= H_5(r)$ to be independent of $z$.
The associated dilaton and the 
antisymmetric tensor field strength are given by
\beqa
e^{-2 \phi} = \frac{1}{H_5} 
\;\;\;,\;\;\; H_{z \theta \phi} = - \partial_{\theta} B_{z \phi} = 
r^2 \sin \theta \, \partial_r H_5' \;\;\;,
\label{nondfb}
\eeqa 
where 
\beqa f = 1 - \frac{\mu}{r} \;\;,\;\; 
H_5 = 1 + \frac{Q_5}{r} \;\;,\;\;  H_5' = 1 + \frac{\alpha_5 Q_5}{r} \;\;,\,\,
Q_5 = \mu \; \sinh^2 \delta_5
\;\;, \;\; 
\alpha_5 = \coth \delta_5 \;.  \eeqa 
The extremal limit is obtained by sending $\mu \rightarrow 0, \delta_5
\rightarrow \infty$ in such a way that the charge $Q_5$ is kept fixed.

We now dualize 
the solution (\ref{nonfiveb}) and (\ref{nondfb}) 
with respect to the isometry direction $z$.
The resulting dual solution
is described by
\beqa
ds^2_{10} &=& \left( 
- f  dt^2 + dx_1^2 + \dots + dx_5^2 \right)
+ H_{5} \left( f^{-1} dr^2  + r^2 \sin^2 \theta d \phi^2 
+ r^2 d \theta^2\right) \nonumber\\
&+& 
\frac{1}{H_{5}} (dz + A_{\phi} d \phi)^2 \;\;\;,\;\;\;
F_{\phi \theta} = - \partial_{\theta} A_{\phi} = r^2 \sin \theta \, \partial_r
H_5' \;\;,\;\; e^{- 2 {\tilde \phi}} = 1 \;.
\label{nonkk}
\eeqa
The field strength of the dual antisymmetric tensor field vanishes.  The line
element (\ref{nonkk}) describes a non-extremal Kaluza-Klein monopole
solution.

Using (\ref{nonkk}), the line element for a non-extremal intersection
of a string, a five-brane, a Kaluza-Klein monopole
and a wave in ten dimensions can be written down according to the rules
given in \cite{cvts}.  In the string frame, it is given by $ds^2_{10} 
= dx_1^2 + \dots + dx_4^2  + ds_6^2$, where $ds_6^2$ denotes the
six-dimensional line element
\beqa
ds^2_{6} &=&  \frac{1}{H_1} \left( 
- H_0^{-1} f  dt^2 + H_0 ( dy + ( H_0'^{-1} -1) dt)^2 \right)
+
H_5 H_{KK} f^{-1} dr^2  + H_5 H_{KK} r^2 d \Omega_2^2 \nonumber\\
&+& 
\frac{H_5}{H_{KK}} (dz + A_{\phi} d \phi)^2 \;\;\;,\;\;\;
 d \Omega_2^2 = \sin^2 \theta d \phi^2 + d \theta^2 \;\;\;,\;\;\;
\label{linesix}
\eeqa
and where
\beqa
f &=& 1 - \frac{\mu}{r} \;\;\;,\;\;\; 
H_i = 1 + \frac{Q_i}{r} \;\;\;,\;\;\; Q_i = \mu \; \sinh^2 \delta_i
\;\;\;,\;\;\;  i = 0,1,5,KK \;\;\;,\;\;\; \nonumber\\
H_0'^{-1}  &=& 1 - \frac{ \mu \; \sinh \delta_0 \; \cosh \delta_0}{r} H_0^{-1} 
\;\;\;,\;\;\; 
H_1'^{-1}  = 1 - \frac{ \mu \; \sinh \delta_1 \; \cosh \delta_1}{r} H_1^{-1} 
\;\;\;,\;\;\; \nonumber\\
H_5' &=& 1 + \frac{\alpha_5 Q_5}{r} \;\;,\;\; 
H_{KK}' = 
1 + \frac{\alpha_{KK}  Q_{KK}}{r} \;\;,\;\; 
\alpha_5 =  \coth \delta_5 \;\;,\;\;
\alpha_{KK} =  \coth \delta_{KK} \;\;,\;\; \nonumber\\
F_{\phi \theta } &=& - \partial_{\theta} A_{\phi}  = r^2 \sin \theta 
\, \partial_r H_{KK}' \;\;\;. 
\label{hq}
\eeqa
Here, the functions $H_i$ ($i=0,1,5,KK$)
denote harmonic functions associated with the wave, the string, the
five-brane and the Kaluza-Klein monopole, respectively.
The six-dimensional dilaton and the 
antisymmetric tensor field are given by \cite{cvts}
\beqa
e^{-2 \phi} = \frac{H_1}{H_5} \;\;\;,\;\;\; B_{ty} = \frac{1}{H_1'} + \, 
{\rm const}
\;\;\;,\;\;\; H_{z \theta \phi} = 
r^2 \sin \theta \, \partial_r H_5' \;\;\;.
\label {dilaten}
\eeqa
Dimensionally 
reducing the six-dimensional line element (\ref{linesix}) to four dimensions
in the $y, z$ directions
and going to the Einstein frame yields the four-dimensional metric
\beqa
ds^2_4 = - \frac{f}{\sqrt{H_0 H_1 H_5 H_{KK}}} \, dt^2 + f^{-1} \;
\sqrt{H_0 H_1 H_5 H_{KK}} \, dr^2 
 + \sqrt{H_0 H_1 H_5 H_{KK}} \, r^2 d \Omega_2^2
\;\;\;.
\label{fourein}
\eeqa
The
internal metric component fields are given by $G_{yy} = H_0/H_1$ and
$G_{zz} = H_5/H_{KK}$.
The four-dimensional dilaton reads
$e^{-2 \phi} = \sqrt{\frac{H_{0}H_1}{H_5 H_{KK}}}$.
This solution
describes a non-extremal four-dimensional black-hole.
The outer horizon is located at $r = \mu$, where $H_i (r = \mu) = \cosh^2 
\delta_i$.  The inner horizon is located at $r=0$.
The Bekenstein-Hawking entropy is
computed to be 
\beqa
{\cal S} = \frac{A}{4G_4} = \frac{\pi}{G_4} \mu^2 \; \cosh \delta_0 \; 
\cosh \delta_1 
 \; \cosh \delta_5 \; \cosh \delta_{KK} \;\;\;.
\label{entrofour}
\eeqa
In the extremal limit $\mu \rightarrow 0, \delta_i \rightarrow \infty \,
(i=0,1,5,KK),$ the Bekenstein-Hawking entropy is given by 
${\cal S} = \frac{\pi}{G_4} \sqrt{Q_0 Q_1 Q_5 Q_{KK}}$ 
\cite{cveyou,cvtsey}.

Let us now consider the four-dimensional line element (\ref{fourein}) in the 
string frame, where it reads
\beqa
ds^2_4 = - \frac{f}{H_0 H_1 } \, dt^2 + f^{-1} \;
 H_5 H_{KK} \, dr^2 + H_5 H_{KK} \, r^2 d \Omega_2^2
\;\;\;.
\label{line4dstring}
\eeqa
We now set $H_0 = H_1$.  Then 
\beqa
ds^2_4 = - \frac{f}{H_1^2} \, dt^2 + f^{-1} \;
  H_5 H_{KK} \, dr^2 
 + H_5 H_{KK} \, r^2 d \Omega_2^2
\;\;\;,\;\;\; e^{-2 \phi} = \frac{H_1}{\sqrt{H_5 H_{KK}}} \;\;\;,
\label{linett}
\eeqa
and the
internal component $G_{yy}$ of the metric is now constant,
$G_{yy} = 1$.  Below we show that 
the line element (\ref{linett}) can be related to the
line element of a two-dimensional black-hole provided that
the harmonic functions $H_5$ and $H_{KK}$ are taken to be 
\beqa
H_5 = \frac{{\cal Q}_5}{r} 
\;\;\;,\;\;\; H_{KK} = \frac{{\cal Q}_{KK}}{r}  \;\;\;,\;\;\;
\label{dualmh}
\eeqa
where the charges ${\cal Q}_5$ and  
${\cal Q}_{KK}$ will turn out to be different from 
the charges $Q_5$ and $Q_{KK}$ introduced in (\ref{hq}).  
Thus, relating the line element (\ref{linett})
to the one of a two-dimensional black-hole requires a change in 
 the asymptotic
behaviour of the magnetic harmonic functions $H_5$ and $H_{KK}$
from the one specified in (\ref{hq}).
Note that with the choice (\ref{dualmh}) the internal metric component
$G_{zz} = H_5/H_{KK}$ is also constant.  Thus, the only
non-trivial scalar field is the dilaton.  Also, with the choice 
(\ref{dualmh}), the line element (\ref{linett}) describes the product
of two two-dimensional spaces, one of which is a two-sphere with 
radius $\sqrt{ {\cal Q}_5 {\cal Q_{KK}}}$.  Each of these two-dimensional
geometries supports a $U(1)$ gauge field.

Below we will show that 
the entropies of the four and the two-dimensional black-holes
match provided that
\beqa
{\cal Q}_5 &=& \alpha_5^2  \; Q_5 = \mu \; \cosh^2 \delta_5 \;\;\;,\;\;\; 
\nonumber\\
{\cal Q}_{KK} &=& \alpha_{KK}^2  \; Q_{KK} = \mu \; \cosh^2 \delta_{KK}
\;\;\;.
\label{nq}
\eeqa
We will furthermore show that (\ref{nq}) is implied
by a sequence of $S$
and $T$-$S$-$T$ duality transformations which brings 
the harmonic functions $H_5$ and $H_{KK}$, given in (\ref{hq}),
into the form (\ref{dualmh}).

Before working out this sequence of duality transformations
we first relate
the solution (\ref{linett}), (\ref{dualmh}) 
to the two-dimensional
heterotic black-hole solution of McGuigan, Nappi and Yost
\cite{guinayo}, as follows.  The latter is described by
\beqa
ds^2_2 &=& - ( 1 - 2m e^{-Qx} + q^2 e^{-2Qx}) dt^2 + 
( 1 - 2m e^{-Qx} + q^2 e^{-2Qx})^{-1} dx^2 \;\;\;, \nonumber\\
e^{-2(\phi - \phi_0)} &=& e^{Qx} \;\;\;,\;\;\;
F_{tx} = \sqrt{2} Q q e^{-Qx} \;\;\;.
\label{bh2}
\eeqa
Here $m$ and $q$ are constants related to the mass 
and to the electric charge of the solution, with $m > 0$ and
$m^2 \geq q^2$.  $Q$ is a positive constant which determines
the associated 
central charge deficit $c= Q^2$.
The asymptotic flat region corresponds to $x = \infty$,
whereas the curvature singularity is at $x = - \infty$.  By changing the
spatial variable to $y = e^{-Qx}$, the solution (\ref{bh2}) becomes
\beqa
ds^2_2 &=& - q^2 (y-y_1)(y-y_2) \, dt^2 + \frac{1}{q^2 Q^2} \frac{dy^2}
{y^2 (y - y_1) (y-y_2)} \;\;\;, \nonumber\\
e^{-2(\phi - \phi_0)} &=& {1 \over y}  \;\;\;,\;\;\;
A_{t} = \sqrt{2} q y \;\;\;.
\label{twobh}
\eeqa
The asymptotic flat region is now at $y=0$, the curvature singularity
is at $y = \infty$, while the two horizons are at $y_{1,2} = q^{-2}
( m \pm \sqrt{m^2 - q^2})$.  
By setting
\beqa
y = \frac{\mu + 2 Q_1}{2m} 
\; \frac{1}{r + Q_1} \;\;\;,
\label{yr}
\eeqa
the solution (\ref{twobh}) turns into the 
two-dimensional $(t,r)$ part of (\ref{linett}), provided that
the parameters $m,q$ and $Q$ 
appearing in (\ref{twobh}) and in (\ref{yr})
are related to the charges $Q_1,{\cal Q}_5$ and 
${\cal Q}_{KK}$
as follows: 
\beqa
Q^2 = \frac{1}{{\cal Q}_5 {\cal Q}_{KK}} \;\;\;,\;\;\;
m e^{-2 \phi_0} = \frac{\mu + 2 Q_1}{2 \sqrt{{\cal Q}_5 
{\cal Q}_{KK}}} \;\;\;,\;\;\;
\frac{m}{q} = \frac{\mu + 2 Q_1}{2 \sqrt{Q_1(Q_1 + \mu)}} \;\;\;.
\label{param}
\eeqa
In the limit $\mu =0$ we get back the extremal two-dimensional
black-hole with $m=q$.

The entropy of the charged non-extremal 
two-dimensional black-hole (\ref{bh2})
was computed in \cite{guinayo,napasq} using thermodynamic methods.
It is given by
\beqa
{\cal S} = \frac{1}{4  G_2} e^{-2 \phi_0} ( m + \sqrt{m^2 - q^2} )\;\;\;.
\label{entrotwo}
\eeqa
Here, $G_2$ denotes the two-dimensional Newton's constant, which was set
to $G_2 = 1/(16 \pi)$ in \cite{guinayo,napasq}.  Inspection of (\ref{linett})
shows that $G_2$ is related to 
Newton's constant $G_4$ in four dimensions by
\beqa
\frac{1}{G_2} = \frac{V}{G_4} \;\;\;,\;\;\; V = 4 \pi {\cal Q}_5 
{\cal Q}_{KK} \;\;\;,
\label{g2g4}
\eeqa
where $V$ denotes the area of the two-sphere with radius $\sqrt{
{\cal Q}_5 {\cal Q}_{KK}}$.
Inserting (\ref{g2g4}) as well as (\ref{param}) into the
entropy formula (\ref{entrotwo}) yields 
\beqa
{\cal S} = \frac{\pi}{G_4} \mu \; \cosh^2 \delta_1  \sqrt{{\cal Q}_5 
{\cal Q}_{KK}} \;\;\;,
\;\;\;,
\label{nonentro}
\eeqa
which is in agreement with the macroscopic formula (\ref{entrofour}) of the 
four-dimensional black-hole provided that (\ref{nq}) holds.

We will now show that there is a sequence of $S$ and $T$-$S$-$T$
duality transformations which can be used to bring
$H_5$ and $H_{KK}$ into the form (\ref{dualmh}) and (\ref{nq}):
\beqa
H_5 = 1 + \frac{Q_5}{r} \rightarrow \frac{{\cal Q}_5}{r} = H_5 -f \;\;\;,\;\;\;
H_{KK} = 1 + \frac{Q_{KK}}{r} \rightarrow \frac{{\cal Q}_{KK}}{r} 
= H_{KK} -f \;\;\;.\;\;\;
\label{hhprime}
\eeqa
We note here that in the extremal case the transformation (\ref{hhprime})
amounts to shifting away the constant part in $H_5$ and $H_{KK}$.
In order to implement the above sequence of duality transformations, 
we first 
dimensionally reduce
the six-dimensional line element
(\ref{linesix}) in the $t, y$ directions, that is
we compactify it on a 
two-torus with Lorentzian signature. This is necessary,
because we will make use of $S$-duality, which
is a symmetry of the heterotic string
in four dimensions.
Note that compactifications on Lorentzian tori as well as
T-duality transformations along time-like directions have recently been
employed in the study of extensions
of U-duality groups in M-theory,
in the construction of Euclidean supergravity theories in various
dimensions as well as in the study of the 
relation of $D$-instantons to other $p$-branes
\cite{Udualgrp,HulJul,Cremetal,Dinst}.

The resulting four-dimensional metric 
in the Einstein frame and the four-dimensional dilaton read
\beqa
ds^2_4 =  \sqrt{f} \left(
f^{-1} H_{KK} dr^2 + H_{KK} \, r^2 d \Omega_2^2
+ \frac{1}{H_{KK}} (dz + A_{\phi} d \phi)^2 \right) \;\;\;,\;\;\;
e^{-2\phi} = \frac{\sqrt{f}}{H_5} \;\;\;.\;\;\;
\label{lineeu}
\eeqa
The antisymmetric 
tensor field $H_{z \theta \phi} 
=  r^2 \sin \theta \, \partial_r H_5' $ 
can be dualized
to an axion field $b^{(5)}$ using 
\beqa
\partial_{\mu} b^{(5)} = \frac{1}{6} e^{- 4 \phi} \sqrt{\det G^{(E)}}
\epsilon_{\mu \nu \rho \sigma} H^{\nu \rho \sigma} \;\;\;.
\label{bh}
\eeqa
We find that 
\beqa
 b^{(5)} = \frac{\alpha_5}{H_5} \;\;\;,\;\;\; \alpha_5 = \coth \delta_5 \;\;\;,
\eeqa
where we have chosen the integration constant in $b^{(5)}$
to be zero.
The scalar fields $b^{(5)}$ 
and $ \phi$ parametrize a coset $SL(2,R)/SO(1,1)$.
It is convenient to introduce the linear combinations \cite{bakas}
\beqa
S_{\pm} = b^{(5)} \pm e^{-2\phi} = \frac{\alpha_5 \pm \sqrt{f}}{H_5} \;\;\;.
\eeqa
$S_{\pm}$ can undergo $SL(2,R)$ transformations
\beqa
S_{\pm} \rightarrow S_{\pm}' = \frac{A S_{\pm} + B}{C S_{\pm} + D} 
\;\;\;,\;\;\;
 AD -BC =1  \;\;\;.
\label{sl2r}
\eeqa
Choosing 
\beqa
A = \frac{1}{\alpha_5} \;\;\;,\;\;\; B = 0 \;\;\;,\;\;\;
C = - \frac{\alpha_5}{1 + \alpha_5} \;\;\;,\;\;\; D = \alpha_5 \;\;\;
\label{dtr}
\eeqa
yields 
\beqa
S_{\pm}' =  b'^{(5)} \pm e^{-2\phi'} = \frac{1 \pm \sqrt{f}}{H_5 -f}\;\;\;,
\eeqa
that is
\beqa
 b'^{(5)} = \frac{1}{H_5 -f} \;\;\;,\;\;\; 
 e^{-2\phi'} =  \frac{ \sqrt{f}}{H_5 -f} \;\;\;.
\eeqa
Thus, the $SL(2,R)$ transformation (\ref{dtr}) precisely results in
the transformation $H_5 \rightarrow H_5 - f$
as in (\ref{hhprime}).
The transformed antisymmetric 
tensor field strength $H'_{z \theta \phi}$ reads
$ H'_{z \theta \phi}
=   r^2 \sin \theta \,  \partial_r (H_5 -f) $. 
In the extremal limit $ \mu \rightarrow 0, \coth \delta_5 \rightarrow 1$,
the parameters of the 
$SL(2,R)$ transformation are given by $A = D = 1, C= - 1/2$.

Next we will show that we can similarly modify the harmonic function
$H_{KK}$ without changing $H_5$. To do so 
we proceed with the Buscher-dualization of the space-time
metric (\ref{lineeu}).  This needs to be done in the string frame.
In the string frame the metric reads
\beqa
ds^2_4 = f^{-1}
H_5 H_{KK} dr^2 + H_5 H_{KK} r^2 d \Omega_2^2 
+ \frac{H_5}{H_{KK}} (dz + A_{\phi} d \phi)^2 \;\;\;,
\label{mstring}
\eeqa
where $F_{\phi \theta } = - \partial_{\theta} A_{\phi}  = r^2 \sin \theta
\partial_r H_{KK}'$.  We note that the reduction of the six-dimensional
line element (\ref{linesix}) in the $t,y$ direction does not lead to
abelian gauge fields in four dimensions.  Moreover, the
components of the six-dimensional metric and of the antisymmetric tensor field
in the $t,y$ directions, which become moduli in four dimensions, do not
depend on $H_5$ and on $H_{KK}$.  Only 
the metric, the antisymmetric
tensor field and the dilaton in four dimensions 
depend on the
magnetic harmonic functions $H_5$ and $H_{KK}$.  Inspection of the 
equations of motion given in \cite{sen} shows that
due to the absence
of gauge fields in four dimensions, the equations of motion for
the moduli decouple from the equations of motion for the four-dimensional
metric, axion and dilaton system.  Thus we may Buscher-dualize 
the metric, axion and dilaton system using (\ref{buscher}).  Dualizing
with respect to the compact direction $z$ yields
\beqa
{\tilde G}_{zz} &=& \frac{H_{KK}}{H_5} \;\;\;,\;\;\; 
{\tilde G}_{zi} = H_{KK} \frac{B_{zi}^{(5)}}{H_5} \;\;\;,\;\;\;
{\tilde G}_{rr} = f^{-1} H_5 H_{KK} \;\;\;,\;\;\; \nonumber\\
{\tilde G}_{lp} &=& H_{KK} \left( H_5 g_{lp} + 
\frac{B_{zl}^{(5)}B_{zp}^{(5)}}{H_5} \right) \;\;\;,\;\;\;
e^{-2 {\tilde \phi}} = \frac{\sqrt{f}}{H_{KK}} \;\;\;,
\label{dualm}
\eeqa
where $i = r, \phi, \theta$ and where
$l,p = \theta, \phi$.  Here $g_{lp}$ denotes the metric of a 
unit two-sphere and
$B_{zi}^{(5)}$ denotes the $B$-field associated with the axion
$b^{(5)}$ introduced in (\ref{bh}).  
$B_{zi}^{(5)}$ only depends on $H_5'$, not on $H_{KK}$.
The dual antisymmetric tensor field reads
\beqa
{\tilde B}_{zi} = A_i \;\;\;,\;\;\; {\tilde B}_{ij} = B_{ij}^{(5)} + A_i 
B_{zj}^{(5)} - A_j B_{zi}^{(5)} \;\;\;.
\eeqa
Since $B_{ij}^{(5)} = 0$ and
$B_{zi}^{(5)} \propto A_i$, it follows that 
\beqa
{\tilde B}_{ij} = 0 \;\;\;.
\eeqa
Hence the only non-vanishing component of the dual antisymmetric tensor field
is ${\tilde B}_{z\phi} = A_{\phi}$.  We thus see that under 
Buscher-dualization $H_5$ and $H_5'$ become interchanged with $H_{KK}$ and
$H_{KK}'$, respectively.  
This is similar to the exchange of $H_5$ and $H_{KK}$
in the extremal case \cite{fuci,behka}.

We now go back to the Einstein frame.  In the Einstein frame
the dual metric reads
\beqa
ds^2_4 = \sqrt{f} 
H_5  \left( f^{-1} dr^2 + r^2 d\Omega_2^2 \right)
+  \frac{\sqrt{f}}{H_5} \left( dz + B_{z \phi}^{(5)} d \phi \right)^2
\;\;\;.
\label{dualmet}
\eeqa
Note that it is independent of $H_{KK}$.  All the dependence
of $H_{KK}$ is contained in the dual fields
\beqa
{\tilde S}_{\pm} = {\tilde b} \pm e^{- 2 \tilde {\phi}} = 
\frac{\alpha_{KK} \pm \sqrt{f}}{H_{KK}} \;\;\;,\;\;\; \alpha_{KK} = \coth
\delta_{KK} \;\;\;.
\eeqa
Thus, as in (\ref{dtr}), we can use an $SL(2,R)$ transformation
with parameters 
\beqa
A = \frac{1}{\alpha_{KK}} \;\;\;,\;\;\; B = 0 \;\;\;,\;\;\;
C = - \frac{\alpha_{KK}}{1 + \alpha_{KK}} 
\;\;\;,\;\;\; D = \alpha_{KK} \;\;\;
\label{dtrk}
\eeqa
to transform ${\tilde S}_{\pm}$ into
\beqa
{\tilde S}_{\pm}' =  
{\tilde b}' \pm e^{-2 {\tilde \phi}'} 
= \frac{1 \pm \sqrt{f}}{H_{KK} -f}\;\;\;,
\eeqa
that is
\beqa
 {\tilde b}' = \frac{1}{H_{KK} -f} \;\;\;,\;\;\; 
 e^{-2 {\tilde \phi}} =  \frac{ \sqrt{f}}{H_{KK} -f} \;\;\;.
\eeqa
Thus, the $SL(2,R)$ transformation (\ref{dtrk}) precisely results in
the transformation $H_{KK} \rightarrow H_{KK} - f$
as in (\ref{hhprime}).
The transformed antisymmetric 
tensor field strength ${\tilde H}'_{z \theta \phi}$ reads
$ {\tilde H}'_{z \theta \phi}
=  r^2 \sin \theta \partial_r (H_{KK} -f) $.

In the string frame, the dualized metric is now given by
\beqa
ds^2_4 = (H_{KK} -f) 
H_5  \left( f^{-1} dr^2 + r^2 d\Omega_2^2 \right)
+  \frac{(H_{KK} -f)}{H_5} \left( dz + B_{z \phi}^{(5)} d \phi \right)^2
\;\;\;.
\eeqa
Then, by Buscher-dualizing again with respect
to $z$ we get back the metric 
(\ref{mstring}) as well as the original dilaton $e^{- 2 \phi} = 
\frac{\sqrt{f}}{H_5}$, 
but now with $H_{KK}$ and  $H'_{KK}$ replaced by $H_{KK} -f$.

Hence we conclude that we can acomplish (\ref{hhprime}) by the
sequence of $S$ and $T$-$S$-$T$ transformations described above.
The six-dimensional line elements (\ref{linesix}) with $H_5$ and
$H_{KK}$ given either as in (\ref{hq}) or as in (\ref{hhprime}) are thus
related by duality transformations.

In the extremal case, equations 
(\ref{param}), (\ref{g2g4}) and (\ref{nonentro})
have a straightforward $O(6,22)$ generalisation in the magnetic sector
through the replacement
\beqa
Q_5 Q_{KK} \rightarrow \frac{1}{2} Q_a L_{ab} Q_b \;\;\;,
\eeqa
where $L_{ab}$ denotes the $O(6,22)$ metric given in \cite{sen}.
On the other hand, in the non-extremal case we have ${\cal Q}_{5,KK} 
= Q_{5,KK} + \mu$.  
The corresponding $O(6,22)$ generalisation
should thus be given by 
\beqa
{\cal Q}_5 {\cal Q}_{KK} \rightarrow  \frac{1}{2} Q^T L Q
+ \mu \, \sqrt{Q^T {\cal M} Q} + \mu^2 \;\;\;,
\eeqa
where 
${\cal M} = L M_{\infty} L + L$ \cite{cvegai}.  Here $M$ 
denotes the moduli matrix of \cite{sen} 
and $ M_{\infty}$ its value at spatial infinity.  $ M_{\infty}$
describes the asymptotic value of the moduli fields occuring in the
four-dimensional black-hole solution.

The six-dimensional solution (\ref{linesix}) was obtained 
by a toroidal compactification of a non-extremal intersection in
ten dimensions.  It remains a solution in six dimensions, if instead
of compactifying heterotic string theory on a four-torus $T_4$ we 
compactify it on $K3$.  A subsequent compactification on a two-torus
leads to a heterotic  $N=2$ compactification in four dimensions.
In heterotic $D=4, N=2$ compactifications the heterotic tree-level 
prepotential will in general receive perturbative and non-perturbative
corrections.  These corrections will result in modifications
of the tree-level solutions and hence also of the associated
Bekenstein-Hawking entropy.  That is, if a tree-level black-hole
solution has an entropy of the form (\ref{nonentro}), then 
this particular form will in general not be preserved 
due to the corrections to the tree-level prepotential.  There are, however,
heterotic $N=2$ compactifications for which the tree-level 
prepotential is exact.  An example of such a compactification is
the FHSV model \cite{fhsv,hm}, where the modulus undergoing $SL(2,R)$ 
transformations is proportional to the heterotic dilaton.
Thus, in this model
a black-hole solution with an entropy of the form
(\ref{nonentro}) will be connected
to a two-dimensional black-hole solution of the form (\ref{twobh})
by a sequence of $S$ and $T$-$S$-$T$ duality transformations.

The tree-level form of the higher-order
derivative gravitational coupling function
$F_1$ is known for four-dimensional
$N=2,4$ heterotic string compactifications.  It would be interesting 
to determine its impact 
on the macroscopic entropy of a four-dimensional
black-hole.  One could try to achieve this by studying its dual two-dimensional
version (\ref{bh2}).
For a conformal field theory description of the
two-dimensional black-hole (\ref{bh2}) we refer to 
\cite{steif,horne,john,sfettsey}.
Knowledge of the associated conformal field theory
should allow one to compute the $\alpha'$-corrections to the
space-time metric of the two-dimensional black-hole and also to its entropy.

{\large \bf Acknowledgements}

\smallskip
\noindent
We would like to thank B. de Wit, S. Massar and E. Verlinde for
discussions.
The work of G. L. C. is supported by 
the European 
Commission TMR 
programme ERBFMRX-CT96-0045.
T. M. thanks the theory group of the University of Utrecht
for its hospitality during his stay there, during which part of this 
work was carried out.

\end{document}